# Method to measure Earth missed by Ancient Greeks?

Fabio Falchi
ISTIL- Istituto di Scienza e Tecnologia dell'Inquinamento Luminoso

Abstract:
I describe a simple method to calculate Earth dimensions using only local measurements and observations. I used modern technology (a digital photo camera and Google Earth) but the exact same method can be used without any aid, with naked eye observations and distances measured by walking, and so it was perfectly accessible to Ancient Greek science.

Eratosthenes was the first to calculate the Earth's dimensions in III century B.C. by measuring the altitude of the Sun at the summer solstice in two cities at different latitudes and measuring their distance, resulted to be about 800 km. Another very simple method can calculate the Earth's radius using only local measurements in the order of few kilometres. Similar approaches to the present one were described recently (e.g. in the references), but the way described here was perfectly accessible to the science of Ancient Greeks and apparently missed by them.

During a brief stay in Sardinia I observed that the line of the horizon is evidently very close to you when the altitude of your eyes is few decimetres above the sea surface. I remember noting this using binoculars from the sea shore years earlier, but it is apparent even with naked eye. This observation needs the sea to be very calm, the calmer the better. A small lake on a no-wind day would be perfect. From the beach, i.e. observing with the eyes at 3-4 m above the water level, two rocks of unknown dimensions were visible emerging from the sea horizon. These rocks were visible even observing them with the eyes at about 1 m above the sea level (fig.1, upper photo). I then noted that once you lower your eyes' position to few centimetres above the water, the rocks progressively disappear under the horizon, hiding completely the left one (fig.1, lower photo). It is impressive to see the horizon hiding the rocks while you bend on your knees. This experience gives you the possibility to touch directly and in an immediate way the curvature of our Earth and its finite dimension. I recommend trying this first-hand at the next available occasion and suggesting it to your students as a summer assignment.

As seen in fig.2, once the distance $d$ from the rocks and their height $h$ are known, the Earth radius $R$ can be computed as follows. The right triangle of fig. 2 is square in $O$, where the observer is. The observer's sight is tangent to the water surface, so we have $(R + h)^2 = d^2 + R^2$

from which follows: $R^2 + 2Rh + h^2 = d^2 + R^2$

then: $2Rh + h^2 = d^2$ and subsequently $2Rh = d^2 - h^2$.

Now $d \gg h$, so we can neglect the term $h^2$ to get $2Rh \approx d^2$ that rearranged becomes:

$$R \approx \frac{d^2}{2h} \qquad (1)$$

I didn't go physically to the rocks to measure their height and their distance. Instead, I used a telephoto lens to photograph them. I then calculated the angular height of each pixel, resulted to be 7.9 seconds of arc, by measuring the dimension in pixel of a millimetre ruler photographed at a known distance of 4.00

m. The height of the rock in the photographs was 16 pixels, corresponding to 126 seconds of arc. The distance $d$ from my position to the rocks was measured using Google Earth (fig. 3) and resulted to be 6.0 km. At this distance, 126 second of arc matches 3.7 metres. This was the height $h$ of the left rock, hidden by the horizon once you observe it by grazing the sea surface. With these data in the equation 1 the Earth radius $R$ results to be 4900 km. The difference with the 'true' known value arises probably from the fact that the very small summit of the rock was invisible to the naked eye, while in reality it was still above the horizon as seen examining closely the photographs (figure 1, inset). What was really hidden under the horizon was a rock's portion of about three metres. With $h$=3 m the Earth radius results to be 6000 km. The method illustrated here gives a precision of about 20% that can be increased using additional observations (different rocks, different distances). A source of uncertainty is given by the fact that in the real situation at sea you cannot position the eyes at exactly the sea level. This issue can be solved by slightly increasing the complexity of the geometry by adding a second triangle, equal to the first and symmetric about the axis given by the segment *CO*. In this case the observer has to watch from a height equal to that of the rock's summit, so it is simpler to select a rock emerging about 2 m over the water level. Then he or she has to travel away from the rock, while staying always at 2 m above the sea level, until the rock will be completely hidden below the horizon. In the case of a rock and an observer 1.5 m high, the distance at which the rock disappears under the horizon is about 8.7 km, while for heights of 2 m the distance will be 10 km. Smaller heights are not advisable because we are approaching the angular optical resolution of the human eye. Solutions with different eye and rock heights are possible, but losing most of the elegance in the simplicity of the method.

While I used modern technology to simplify my measurements, the essential parts of the method, including both the geometry and the distance and height measurements, was perfectly accessible by ancient Greeks. This method is surely more direct than the one used by Eratosthenes, which relied on the measurement of a very long distance and on the assumption that the Sun's rays are parallel (i.e. the Sun is very far, compared to the distance between the cities of Syene and Alexandria). A closer Sun could explain the difference that Eratosthenes observed in the Sun's altitude even with a flat Earth. This method also uses a simpler mathematics than the one described elsewhere (e.g. Kibble 2011), not having to use a system of two equations.

Why, then, did the Ancient Greeks miss this simple method to measure Earth dimension? Or, are we the one missing some lost ancient Greek source? Greeks knew that ships progressively disappear under the horizon while going away to the open sea and correctly explained this phenomenon with the curvature of Earth. They also lived in places similar to the one where I made my observations, with plenty of rocks emerging some metres above the sea surface. Maybe they didn't enjoy swimming and staying in the sea, so they missed the chance to find an alternative method to measure Earth radius.

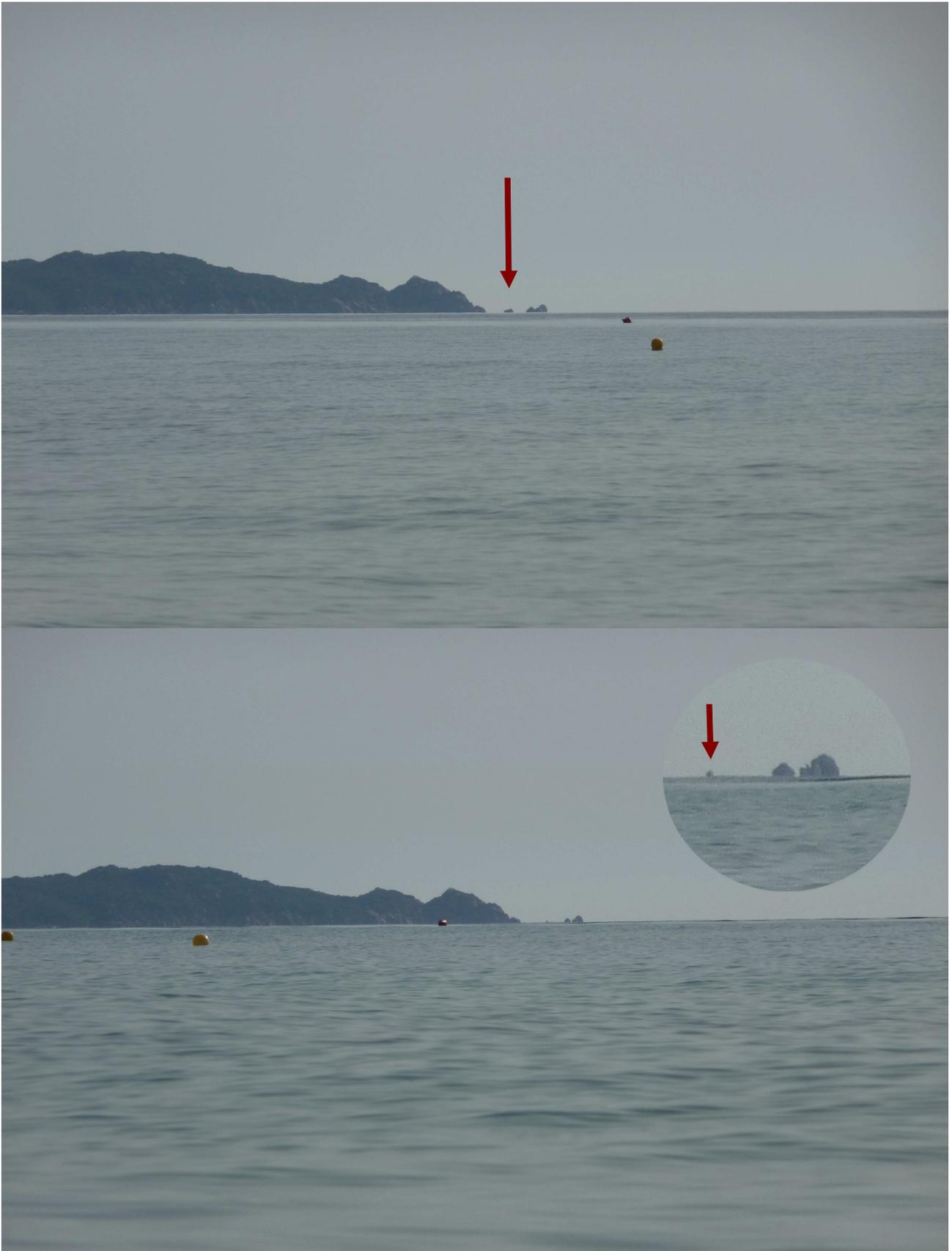

Fig. 1. The left rock disappears to the eyes as seen from few centimetres above the sea level (lower photo), while it is visible observing from 1 m above (arrowed rock, upper photo). Part of the rock is still visible examining carefully the lower photo (inset, short arrow). Observing with eyes only it is possible to get closer to the water surface, while using a camera it is important to be careful not to submerge it, unless it is waterproof.

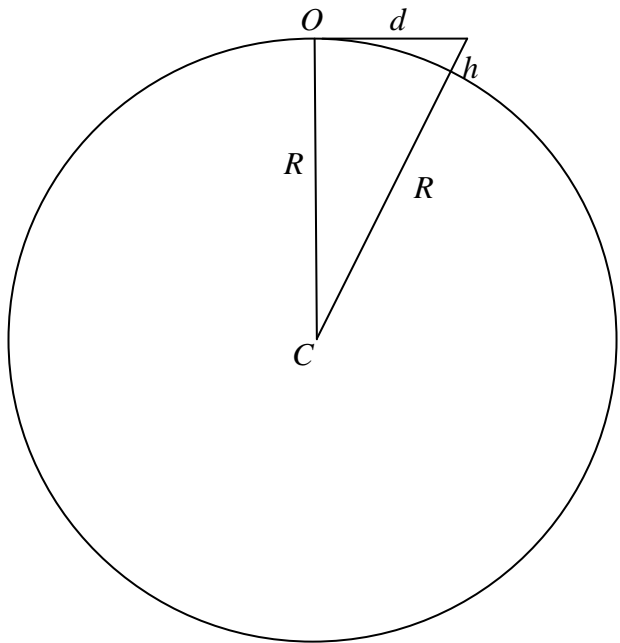

Fig. 2.

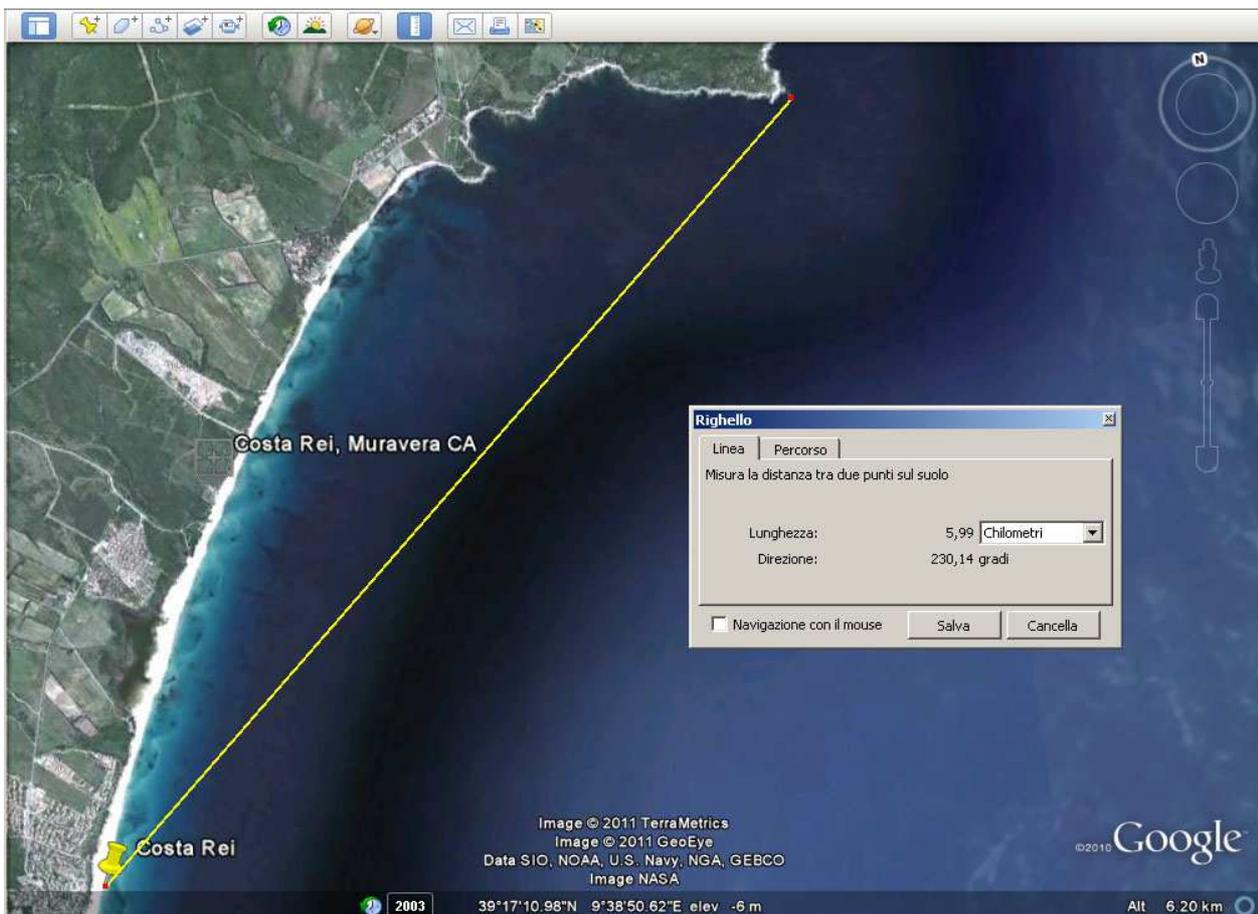

Fig.3 The distance between the observer and the rocks is easily found with Google Earth (here 5.99 km), but it would have been simple even for ancient Greeks to determine the distance of few kilometres.